\documentstyle[11pt,aaspp4]{article}
\received{4 April 2000 }
\accepted{21 June 2000}
\slugcomment{{\bf To be published in the Astrophysical Journal} \\ Vol. 544, 20 Nov. 2000} 

\lefthead{SENGUPTA}
\righthead{JOULE HEATING IN NEUTRON STARS}
\begin{document}

\title{JOULE HEATING IN NEUTRON STARS UNDER STRONG GRAVITATION}

\author{SUJAN SENGUPTA}
\affil{Indian Institute of Astrophysics \\
       Koramangala, Bangalore 560 034, India. E-mail : sujan@iiap.ernet.in}

\begin{abstract}

Considering Joule heating caused by the dissipation of the magnetic field in
the neutron star crust to be an efficient mechanism in maintaining a relatively
high surface temperature in very old neutron stars, the role of general
relativity is investigated. It is found that, although the effect of
space-time curvature produced by the intense gravitational field of the star
slows down the decay rate of the magnetic field, modification of the
initial magnetic field configuration and the initial field strength by
the space-time curvature results in increasing the rate of Joule heating.
 Hence the space-time curvature supports Joule heating in maintaining a relatively high
surface temperature which is consistent with the observational detection.  
\end{abstract}


\section{INTRODUCTION}

Neutron stars cool down mainly by neutrino emission from the inner layers during the
first million years and subsequently by photon emission from the surface.
The rate of photon emission depends on the physical properties of matter inside
the star and on the magnetic field.

Observational inference of the thermal radiation of several old neutron stars in the
X-ray band (Becker \& Trumper, 1997), and in the UV range (Pavlov, Stringfellon
\& Cordova, 1996; Mignani, Caraveo \& Bignami, 1997) indicates much higher surface
temperature as compared to the predictions of the standard cooling models.
Therefore additional heating mechanisms are needed in order to remove the discrepancy
between the detected temperature and the theoretical models.

One of the two possible mechanisms is the heat generated by the frictional
energy of neutron super-fluid with normal matter in the inner crust (Shibazaki \&
Lamb 1989; Umeda et al. 1993) which is independent of the magnetic field. The
other mechanism of additional heating is associated with the ohmic dissipation
of currents (Miralles, Urpin \& Konenkov 1998) which is strongly sensitive to
the configuration and strength of the magnetic field. These authors found that the
 ohmic dissipation produces enough heat to change the thermal evolution of neutron
stars substantially at the late stage although observational data on the magnetic
field evolution of isolated pulsars supports a slow decay rate.

Considering magnetic field configurations which are initially confined to a small
part of the crust and which vanish in the stellar core, several authors 
(Chanmugam \& Sang 1989; Urpin \& Muslimov 1992; Urpin \& Van Riper 1993)
found that the relatively low electrical conductivity of the crustal matter
causes the decay times too short to be of observational interest if the
impurity content is high.
Sengupta (1997, 1998) investigated the
 contribution of space-time curvature on the decay rate
of the crustal magnetic fields in isolated neutron star by assuming a
spherically symmetric stationary gravitational field. It was demonstrated
clearly by Sengupta (1998) that the role of impurity content which increases
the decay rate is suppressed by the effect of space-time curvature. As a result,
even with high impurity content, the decay rate is significantly less at
the late stage of evolution if general relativistic effects are taken
into consideration.

In the present paper, I investigate the effect of space-time curvature produced
by the intense gravitational field of the star on the rate of Joule heating
caused by ohmic dissipation. The basic equations that describe the magnetic
field evolution and the rate of Joule heating under general relativistic framework
are presented in the next section. In section~3 the model adopted in this
investigation is described. The results are discussed in section~4 and the conclusions
are drawn in section~5.

\section{EQUATIONS FOR MAGNETIC FIELD EVOLUTION AND JOULE HEATING }

 Assuming hydrodynamic motions to be negligible and the anisotropy of the electrical
conductivity of the crustal material is small, the induction equation in flat space-time can be
written as:

  \begin{equation}\label{FLAT}
\frac{\partial{\bf B}}{\partial t}=-{\bf \nabla} \times \left(\frac{c^2}{4\pi
\sigma}{\bf \nabla}\times{\bf B}\right),
\end{equation}
where $\sigma$ is the electrical conductivity.

If a stationary gravitational field is taken into account, then using the covariant
form of Maxwell equations and the generalized Ohm's law (Sengupta 1998)
 (neglecting the displacement
current and taking $u^i=0$) the corresponding 
induction equation in curved space-time can be derived as:
\begin{equation}\label{GTR1}
\frac{\partial F_{kj}}{\partial x^0}=
\frac{\partial}{\partial x^k}\left[\frac{c}{4\pi}
\frac{1}{\sqrt{-g}}g_{\mu j}\frac{1}{\sigma u^0}
\frac{\partial}{\partial x^i}(\sqrt{-g}F^{\mu i})\right]
- \frac{\partial}{\partial x^j}\left[\frac{c}{4\pi}\frac{1}{\sqrt{-g}}
g_{\mu k}\frac{1}{\sigma u^0}\frac{\partial}{\partial x^i}(\sqrt{-g}
F^{\mu i})\right],
\end{equation}
where $F_{\mu\nu}$ are the components of the electro-magnetic field tensor,
$J^{\mu}$ are the components of the 
four-current density, $u^{\mu}$ are the components of the four velocity of
the fluid, $g_{\mu\nu}$ are the components of space-time metric that
describes the background geometry and $g={\rm det}|g_{\mu\nu}|.$ 
Here and afterwards Latin indices run over spatial co-ordinates only whereas
Greek indices run over both time and space co-ordinates.

For the description of the background geometry I consider the exterior
Schwarzschild metric which is given by
\begin{equation}\label{sch}
ds^2=(1-\frac{2m}{r})c^2dt^2-(1-\frac{2m}{r})^{-1}dr^2-r^2(d\theta^2+
\sin^2\theta d\phi^2),
\end{equation}
where $m=MG/c^2$, $M$ being the total gravitational mass of the core.
The justification for adopting the exterior Schwarzschild metric is provided
in Sengupta (1998). Since,
the crust consists of less than a few percent of the total gravitational mass,
$M$ can be regarded as the total mass of the star.

Using the metric given in equation (\ref{sch}), equation (\ref{GTR1})
can be reduced to 
\begin{equation}\label{GTR1a}
\frac{\partial F_{kj}}{\partial x^0}= \frac{c}{4\pi}
\left[\frac{\partial}{\partial x^k}\left\{
\frac{1}{r^2\sin\theta}g_{lj}\frac{1}{\sigma u^0}
\frac{\partial}{\partial x^i}(r^2\sin\theta F^{li})\right\}-
 \frac{\partial}{\partial x^j}\left\{\frac{1}{r^2\sin\theta }
g_{lk}\frac{1}{\sigma u^0}\frac{\partial}{\partial x^i}(r^2\sin\theta F^{li})
\right\}\right].
\end{equation}

Following the convention, I consider the decay of
a dipolar magnetic field which has axial symmetry so that the vector potential
${\bf A}$ may be written as $(0,0,A_{\phi})$ in spherical polar co-ordinates
 where $A_{\phi}=A(r,\theta,t).$ Since the hydrodynamic motion is negligible
 so $u^i= dx^i/ds = 0$ and the metric gives
$$u^0=(1-\frac{2m}{r})^{-1/2}.$$

Therefore, from equation (\ref{GTR1a}) we obtain using
the definition
$F_{\alpha\beta}=A_{\beta,\alpha}-A_{\alpha,\beta},$
\begin{equation}\label{GTR2}
\frac{\partial A_{\phi}}{\partial t}=\frac{c^2}{4\pi\sigma}(1-
\frac{2m}{r})^{1/2}\sin\theta\left[\frac{\partial}{\partial r}\left\{(1-
\frac{2m}{r})\frac{1}{\sin\theta}\frac{\partial A_{\phi}}{\partial r}
\right\}+\frac{\partial}{\partial \theta}\left(\frac{1}{r^2\sin\theta}
\frac{\partial A_{\phi}}{\partial \theta}\right)\right].
\end{equation}

Choosing
\begin{equation}
A_{\phi}=\frac{f(r,t)}{r}\sin\theta
\end{equation}
for the flat space-time and
\begin{equation}
A_{\phi}=-g(r,t)\sin^2\theta
\end{equation}
for the curved space-time,
where $r$ and $\theta$ are the spherical radius and polar angle
respectively one gets from equation (\ref{FLAT}) and equation (\ref{GTR2}) 
\begin{equation}\label{FLAT1}
\frac{\partial^2f(r,t)}{\partial r^2}-\frac{2}{r^2}f(r,t)=\frac{4\pi 
\sigma}{c^2}\frac{\partial f(r,t)}{\partial t}
\end{equation}
and
\begin{equation}\label{GTR3}
(1-\frac{2m}{r})^{1/2}\left[(1-\frac{2m}{r})\frac{\partial^2g(r,t)}{\partial r^2}
+\frac{2m}{r^2}\frac{\partial g(r,t)}{\partial r}-\frac{2}{r^2}g(r,t)\right]=
\frac{4\pi\sigma}{c^2}\frac{\partial g(r,t)}{\partial t}
\end{equation}
respectively.

The $\phi$ component of the electric current maintaining the dipolar magnetic
field configuration for flat space-time is given by

\begin{equation}
j_{\phi}=-\frac{c}{4\pi}\frac{\sin\theta}{r}\left(\frac{\partial^2 f}{\partial r^2}-
\frac{2 f}{r^2}\right).
\end{equation} 

In Schwarzschild space-time geometry the above quantity can be written as :
\begin{equation}
j_{\phi}=-\frac{c\sin^2\theta}{4\pi}
\left[(1-\frac{2m}{r})\frac{\partial^2g(r,t)}{\partial r^2}
+\frac{2m}{r^2}\frac{\partial g(r,t)}{\partial r}-\frac{2}{r^2}g(r,t)\right].
\end{equation}

When transfered to a Locally Lorentz frame the above equation takes the following form: 
\begin{equation}
j_{\phi}=-\frac{c}{4\pi}\frac{\sin\theta}{r}
\left[(1-\frac{2m}{r})\frac{\partial^2g(r,t)}{\partial r^2}
+\frac{2m}{r^2}\frac{\partial g(r,t)}{\partial r}-\frac{2}{r^2}g(r,t)\right].
\end{equation}

Following the approach of Miralles, Urpin and Konenkov (1998), I neglect non-sphericity
in cooling calculations and use the angle-averaged expression for Joule heating.
The rate of Joule heating $\dot{q}=j^2/\sigma$ in flat space-time is written as :
\begin{equation}
\dot{q}=\frac{c^2}{24\pi^2 r^2\sigma}\left(\frac{\partial^2 f}{\partial r^2}-
\frac{2f}{r^2}\right)^2.
\end{equation}

Normalizing the function $f(r,t)$ to its initial value at the surface $f(R,0)$
which is related to the initial field strength at the magnetic equator $B_e$
by the relation $f(R,0)=R^2B_e$, the expression for the rate of Joule heating
in flat space-time can be written as
\begin{equation}
\dot{q}=\frac{c^2R^4B^2_e}{24\pi^2r^2\sigma}\left(\frac{\partial^2F}{\partial r^2}-
\frac{2F}{r^2}\right)^2
\end{equation}
where $F(r,t)=f(r,t)/f(R,0)$.

Similarly in curved space-time :
\begin{equation}
\dot{q}=\frac{c^2}{24\pi^2\sigma r^2}
\left[(1-\frac{2m}{r})\frac{\partial^2g(r,t)}{\partial r^2}
+\frac{2m}{r^2}\frac{\partial g(r,t)}{\partial r}-\frac{2}{r^2}g(r,t)\right]^2.
\end{equation}

The function $g(r,t)$ is normalized to its initial value at the surface by
$G(r,t)=g(r,t)/g(R,0)$ and $g(R,0)$ is related to the initial field strength
at the magnetic equator by the expression :
\begin{eqnarray} 
g(R,0) & = &-\frac{B_eR}{2m}\left[R^2\ln(1-\frac{2m}{r})+2mR+2m^2\right]\times \nonumber \\
& &\left[\frac{R}{m}\ln(1-\frac{2m}{R})+(1-\frac{2m}{R})^{-1}+1\right]^{-1}
\left(1-\frac{2m}{R}\right)^{-1/2}.
\end{eqnarray} 

In the present work I have assumed that the space-time metric within the
crustal region and exterior to the stellar surface is the same. Since there exists
a plasma filled magnetosphere outside the surface of the star, one should have
a time varying dipole magnetic field outside the boundary. However, the electric
conductivity of the surrounding plasma is such that the time variation of the
magnetic field is effectively negligible as compared to that within the crust.
Hence for both relativistic and non-relativistic cases, I impose the usual
boundary conditions as given in Urpin \& Muslimov (1992).

\section {THE MODEL}

 The formalisms adopted
in the present work have been discussed in details by  Miralles, Urpin \&
Konenkov (1998). The evolution of the magnetic field in flat space-time
is discussed extensively by Urpin \& Muslimov (1992) and in curved space-time
by Sengupta (1998). 
In the present investigation, the gravitational mass of the star is taken to be 
$1.4 M_{\odot}$. However, two different configurations for the neutron star model
 are adopted : one with a radius 7.35 km and the other with a radius 11 km. The
first one is obtained if the equation of state of the matter inside the star is soft
and the second one is obtained if it is intermediate or stiff.

If one assumes the initial value of $f(r,t)=f(r)$ at $t=0$ for flat space-time,
then for curved space-time (Wasserman \& Shapiro 1983; Sengupta 1995) 
  \begin{equation}
g(r,0)=g(r)=\frac{3rf(r)}{8m^3}[r^2\ln(1-\frac{2m}{r})+2mr+m^2].
  \end{equation}
This is because of the fact that any given magnetic field configuration in flat
space-time is modified by the curvature of space-time produced by the gravitational
field of the central object. Asymptotically at a large distance $g(r)$ coincides
with $f(r)$.

I have considered the decay of the magnetic field which initially occupies
the surface layers of the crust up to a depth  $x=0.966$ where $x=r/R$, R being
the radius of the star.
 The crustal region is considered to
be extended upto $x=0.875$. It should be noted that for the same value of $x$,
the corresponding density for the two different mass-radius configurations is
different. 
The main aim of the present work is to investigate the effect of space-time
curvature produced by the intense gravitational field of the star to the rate
of Joule heating and hence to the thermal evolution of old neutron stars.

The electrical conductivity within the crust has been calculated following the
approaches of Urpin \& Van Riper (1993).
The net conductivity of the crustal material at a given depth is computed as
\begin{equation}
\sigma=\left(\frac{1}{\sigma_{ph}}+\frac{1}{\sigma_{imp}}\right)^{-1},
\end{equation}
where $\sigma_{ph}$ is the conductivity due to electron-phonon scattering
and $\sigma_{imp}$ is the conductivity due to electron-impurity scattering.
The effect of electron-ion scattering has been neglected
since the region where this effect could be important is sufficiently thin.
$\sigma_{imp}$ is inversely proportional to the impurity parameter $\epsilon$ defined
as
\begin{eqnarray}
\epsilon=\frac{1}{n}\sum_{i}n_i(Z-Z_i)^2,
\end{eqnarray}
where $n$ and $Z$ are the number density and electric charge of background ions
in the crust lattice without impurity, $Z_i$ and $n_i$ are the charge and density
of the $i$-th impurity species. The summation is extended over all species of
impurity. It is worth mentioning that the value of the impurity parameter $\epsilon$ 
for neutron star crust is not known at present. Electron-impurity scattering
becomes more important with increasing density and decreasing temperature. Hence
at the late stage of evolution when the temperature becomes low, the conductivity
is dominated by electron-impurity scattering. 
The impurity parameter $\epsilon$ has been taken as $0.01$ and $0.1$.

\section{RESULTS AND DISCUSSIONS}

The evolution of the surface magnetic field normalized to its initial value
for both the general relativistic and the non-relativistic cases with the standard
cooling model and with the impurity
parameter $\epsilon=0.01$ and $\epsilon=0.1$ are presented in Figure~1 
and in Figure~2 respectively.
\placefigure{fig1}
\placefigure{fig2}

It is shown by Muslimov \& Urpin (1992) that the field behavior is qualitatively
independent of the forms of the initial configurations but the numerical
results differ for various choices of the initial depth penetrated by the
magnetic field.  At a very early stage of evolution when the crustal matter 
is melted in the layers of a maximal current density, the magnetic field
does not decay appreciably. After the outer crust solidifies, significant
decay takes place and after 1 Myr no decay occurs if the impurity content is
zero.

The above scenario for flat space-time is not altered with the inclusion of
general relativistic effects but significant decrease in the numerical
value at the late stage of evolution is found when the effect of space-time curvature is incorporated.
The impurity-electron scattering is
dominant at the late stage of evolution when the crustal temperature is low.
As a consequence the decay rate after $t > 10$ Myr changes appreciably. It should
be mentioned here that at the late stage the  electric conductivity becomes
independent of temperature.

\placefigure{fig3}
\placefigure{fig4}

Figure~3 and Figure~4 show the rate of Joule heating for the flat space-time
and for the curved space-time with $\epsilon=0.01$ and $\epsilon=0.1$ respectively. For both
the cases we notice that the rate of Joule heating increases by almost one order of
magnitude when the stellar radius is 7.35 km. In both the Figure~3 and Figure~4 the results
with the stellar radius R=11 km are also presented in order to demonstrate the
effect of compactness. The rate of heat production is equal to the rate of decrease of the magnetic
energy which is proportional to the square of the magnetic field strength at any
time. Figure~1 and Figure~2 show that the rate of magnetic field decay
decreases substantially when the effect of general relativity is incorporated.
As a result one expects the rate of Joule heating to be less when the
effect of general relativity is incorporated. However, the rate of Joule heating
is very much sensitive to the configuration and strength of the initial
magnetic field. The more is the initial field strength the more is the heat
generated. The space-time curvature produced by the intense gravitational
field of the star changes the geometry and increases the strength of the initial
magnetic field substantially. The combination of the two phenomenon, eg., a slower decay rate
but higher initial field strength, yields 
in an overall increase in the rate of Joule heating by almost one order of magnitude. 
The whole purpose of this investigation is to understand the effect of
strong gravitation of the neutron star. The effect of the different 
equation of states of matter and the effect due to the depth penetrated 
by the initial magnetic field have been investigated in details by
Miralles, Urpin \& Konenkov (1998). 

Following the approach of Miralles, Urpin \& Konenkov (1998), I consider
the surface temperature to follow the equation
\begin{equation}
\dot{Q}=4\pi R^2 \sigma_{sb} T^4_s,
\end{equation} 
where $\sigma_{sb}$ is the Stefan-Boltzmann constant.
This is valid because of the fact that
except for a short initial period of 3 to 10 Myr when the surface 
temperature is very high, approximately all heat released due to the
magnetic field dissipation is emitted from the surface. At the earlier
age the influence of Joule heating is not important. At the late
stage the surface temperature is determined by balancing the Joule heating
integrated over the neutron star volume with the photon luminosity.
It should also be mentioned that at the late stage
when electron-impurity scattering becomes dominant, the electric
conductivity becomes independent of the temperature of the crust.

\placefigure{fig5}
\placefigure{fig6}
\placefigure{fig7}
\placefigure{fig8}

The surface temperature with and without the effect of general relativity is presented in
Figure~5 and Figure~6 for $\epsilon=0.01$ and $\epsilon=0.1$ respectively with
R=7.35 km.
Since the earlier evolution is not effected by
Joule heating the results are shown for the age $t>2.5\times10^6$ yr. For
comparison the surface temperature without Joule heating is also presented.  
Figure~7 and Figure~8 present the same results with R=11 km. 
The results show that for both the cases, with or without the effect of
general relativity, the surface temperature increases if the initial magnetic
field strength is increased.  
Figure~5 to Figure~8 clearly show that although  general
relativistic effect slows down the magnetic field decay rate substantially,
the change in the initial magnetic field configuration and the strength of 
the initial field result into an increase in the surface temperature of the star 
as compared to the result obtained without the effect of general relativity.
Although this increase is not very large, the results clearly indicates that
general relativity supports Joule heating in maintaining a relatively high surface temperature of old
neutron stars which is consistent with observational detection. 

Pavlov, Stringfellow and Cordova (1996) estimated the bolometric luminosities
of three isolated pulsars, B0656+14 , B0950+08 and B1929+10. Becker and Trumper (1997)
presented the upper limit of the bolometric luminosities of several milli-second
pulsars. The temperature of a pulsar can be calculated from the observed
 flux by assuming black-body radiation and keeping in mind the fact that the
temperature as calculated in the surface is higher by a factor of $1+z=
(1-2GM/Rc^2)^{-1/2}$. The surface temperature of all the milli-second pulsars
as presented by Becker and Trumper (1997) is several orders of magnitude
higher than that predicted by the standard cooling models. However, it is
believed that the milli-second pulsars are in accretion stage. Therefore
they should have much higher surface temperature than that of isolated pulsars
and their thermal evolution is different than that of isolated pulsar. It
should be pointed out here that the impurity content doesn't have any significant role
at a high temperature. Hence, unlike isolated pulsars, the general relativistic effects 
 should not be significant for the case of accreting neutron stars even at the
late stage of evolution.

In Figure~5 to Figure~8, I have presented the surface temperature of three isolated pulsars :
B0823+26, B1929+10 and B0950+08 whose bolometric luminosities are estimated
from observations (Pavlov, Stringfellow \& Cordova 1996; Becker \& Trumper 1997).
The surface temperature of three isolated pulsars  which are older than $2.5\times 10^6$ yrs 
have been considered from the above references because
 Joule heating is important only after the age $t > 2.5\times 10^6$ yrs. 
Comparison with the theoretical  models that incorporate Joule heating shows
that the inclusion of general relativistic effects makes the results more
consistent with the observational data. It is worth mentioning that the observational
data provides only the upper limit of the bolometric luminosity owing to the
uncertainties in the distance measure of the pulsars.

\section{CONCLUSIONS}
Considering Joule heating caused by the decay of the crustal magnetic field
in neutron stars to be a potential mechanism in explaining the high
surface temperature detected in many old neutron stars, the effect of
space-time curvature produced by the intense gravitational field of the
star on the thermal evolution is investigated. In spite of the fact that
general relativity slows down the magnetic field decay rate substantially,
the rate of Joule heating increases almost by an order of magnitude because
the initial field configuration and the field strength get modified by
the space-time curvature. As a result, general relativistic effects
support Joule heating in maintaining a high surface temperature  at a
very late stage of evolution of isolated neutron stars. 

\acknowledgments
 I am thankful to the referee for valuable suggestions which have improved
the quality of the paper. Thanks are due to Vinod Krishan and N. K. Rao
for their encouragement.

\clearpage

\figcaption[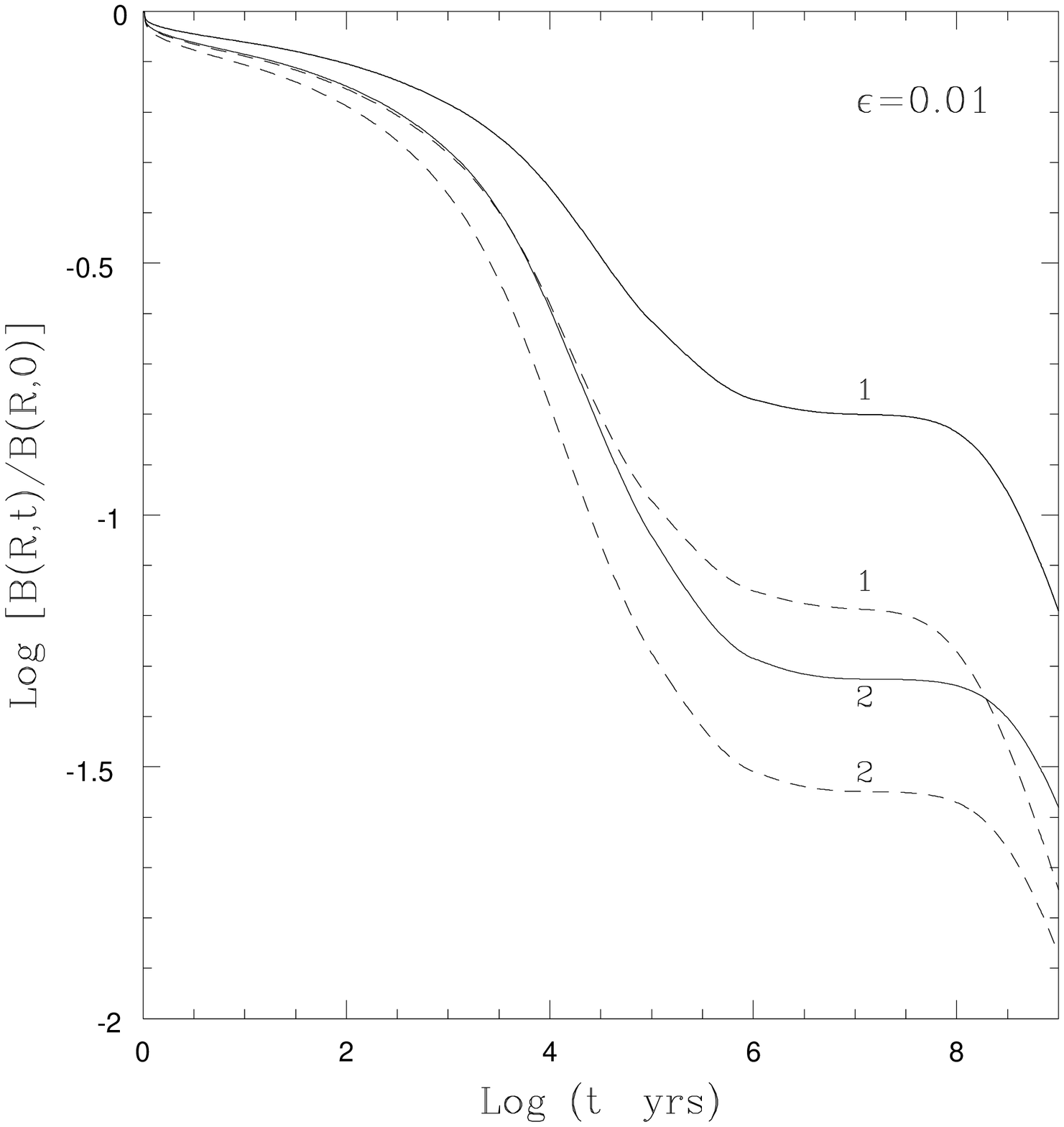]{The evolution of surface magnetic field normalized
to its initial value for flat and curved space-times with the impurity
parameter $\epsilon=0.01$. Solid line represents
the results for curved space-time while broken line represents that for flat
space-time.  Curves 1 represent the results 
for neutron star with mass 1.4 $M_\odot$ and radius 7.35 km while curves 2
represent results for neutron stars with mass 1.4 $M_\odot$ and radius 11 km.
The results are obtained by using the standard cooling model without Joule heating.  
\label{fig1}}

\figcaption[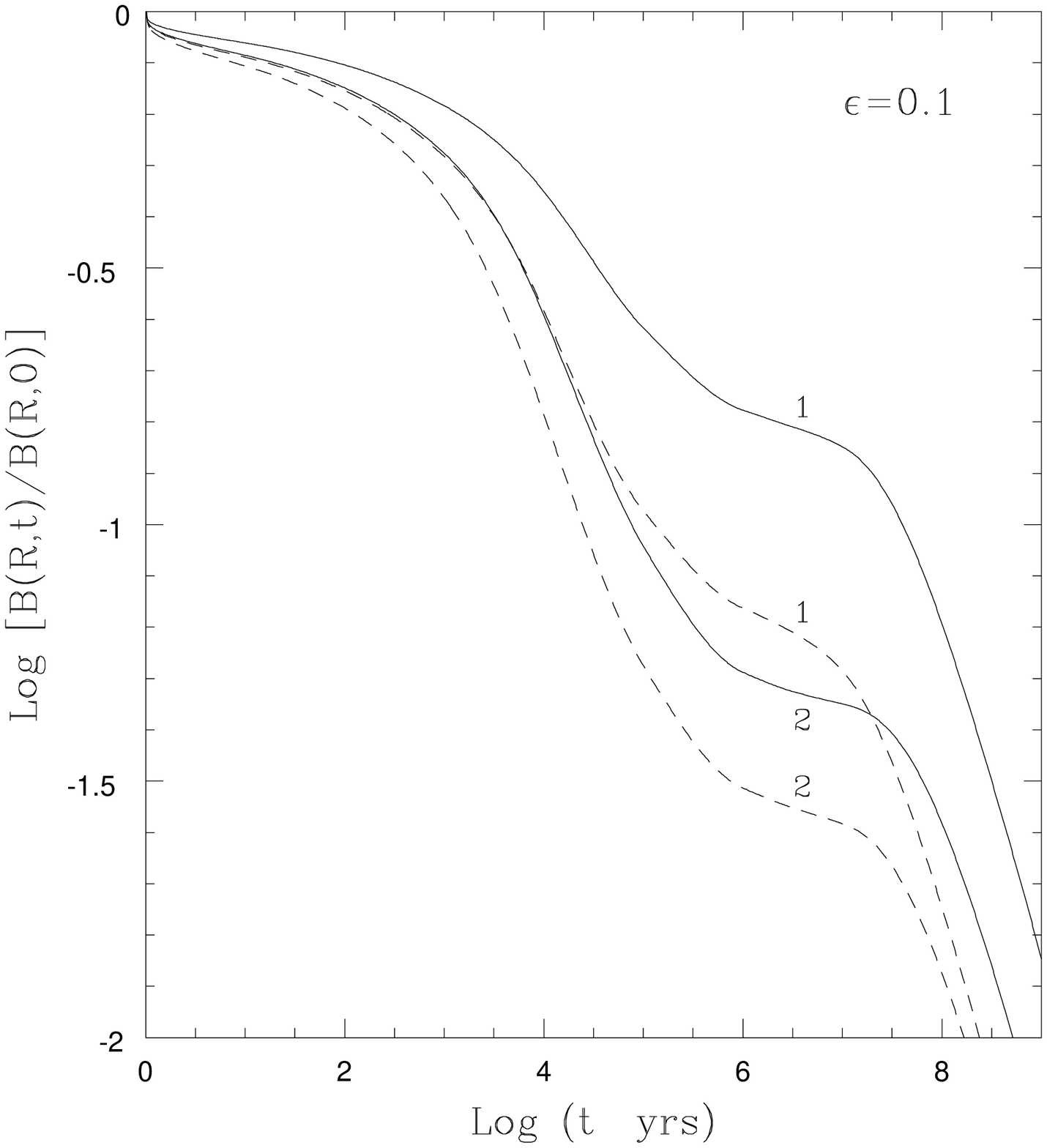]{Same as figure 1 but with $\epsilon=0.1$. \label{fig2}}

\figcaption[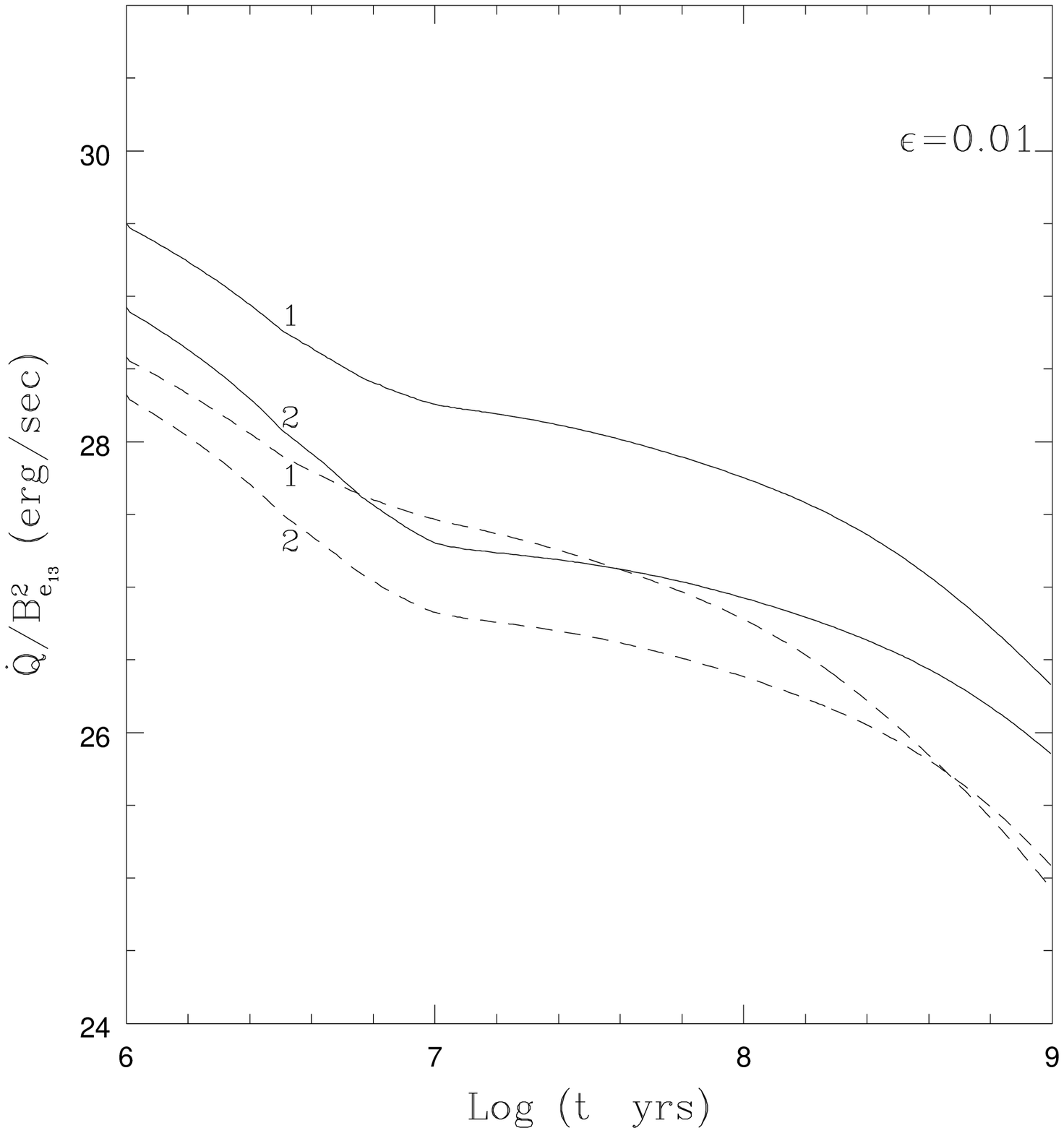] {Temporal dependence of the rate of Joule heating
integrated over the volume of the neutron star and normalized to
$B_e/10^{13}$ G with $\epsilon=0.01$.
 Solid line represents
the results for curved space-time while broken line represents that for flat
space-time.
 Curves 1 represent the results 
for neutron star with mass 1.4 $M_\odot$ and radius 7.35 km while curves 2
represent results for neutron stars with mass 1.4 $M_\odot$ and radius
11 km. \label{fig3}}

\figcaption[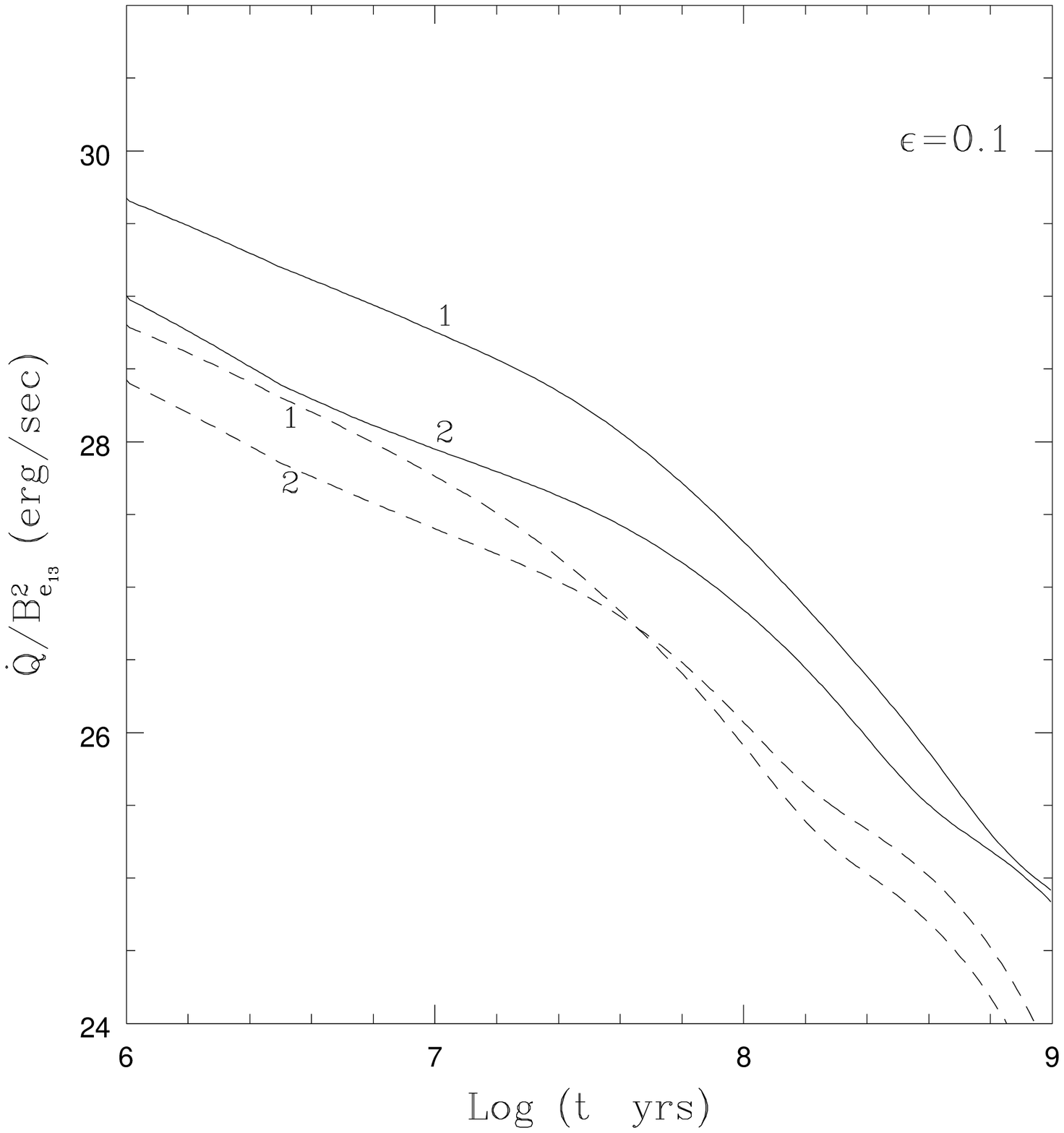]{Same as figure 3 but with $\epsilon=0.1$.
\label{fig4}}

\figcaption[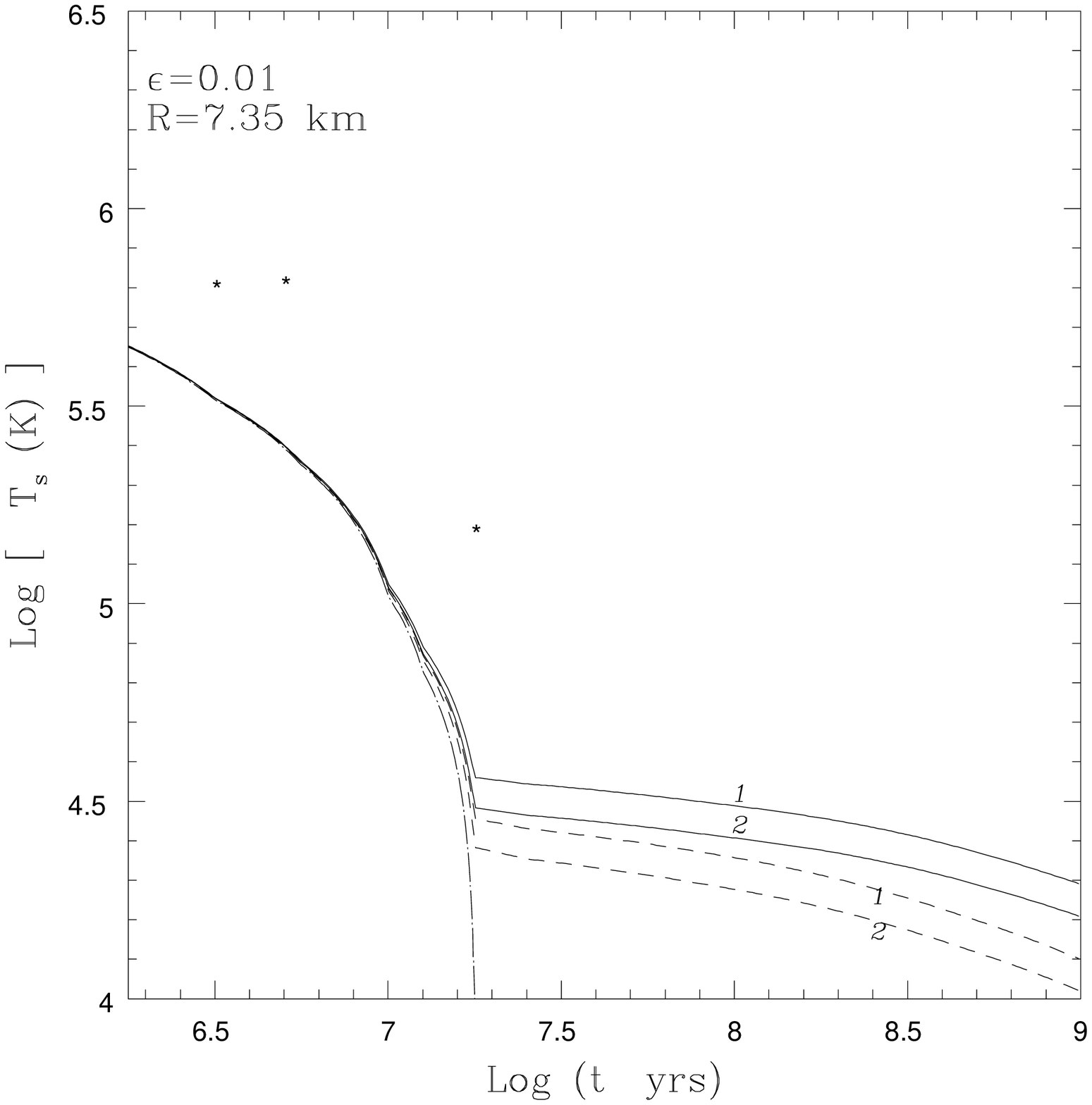] {Surface temperature of neutron star with and without
Joule heating. $\epsilon=0.01$.
 Solid line represents
the results for curved space-time, broken line represents that for flat
space-time and dash-dot line represents the surface temperature without Joule heating.
Curves 1 represents the surface temperature with $B_e=3.0\times 10^{13}$ G
and Curves 2 represents the surface temperature with
$B_e=1.5\times 10^{13}$ G. Stars represent the surface temperature of
isolated neutron stars inferred from observation.  \label{fig5}}

\figcaption[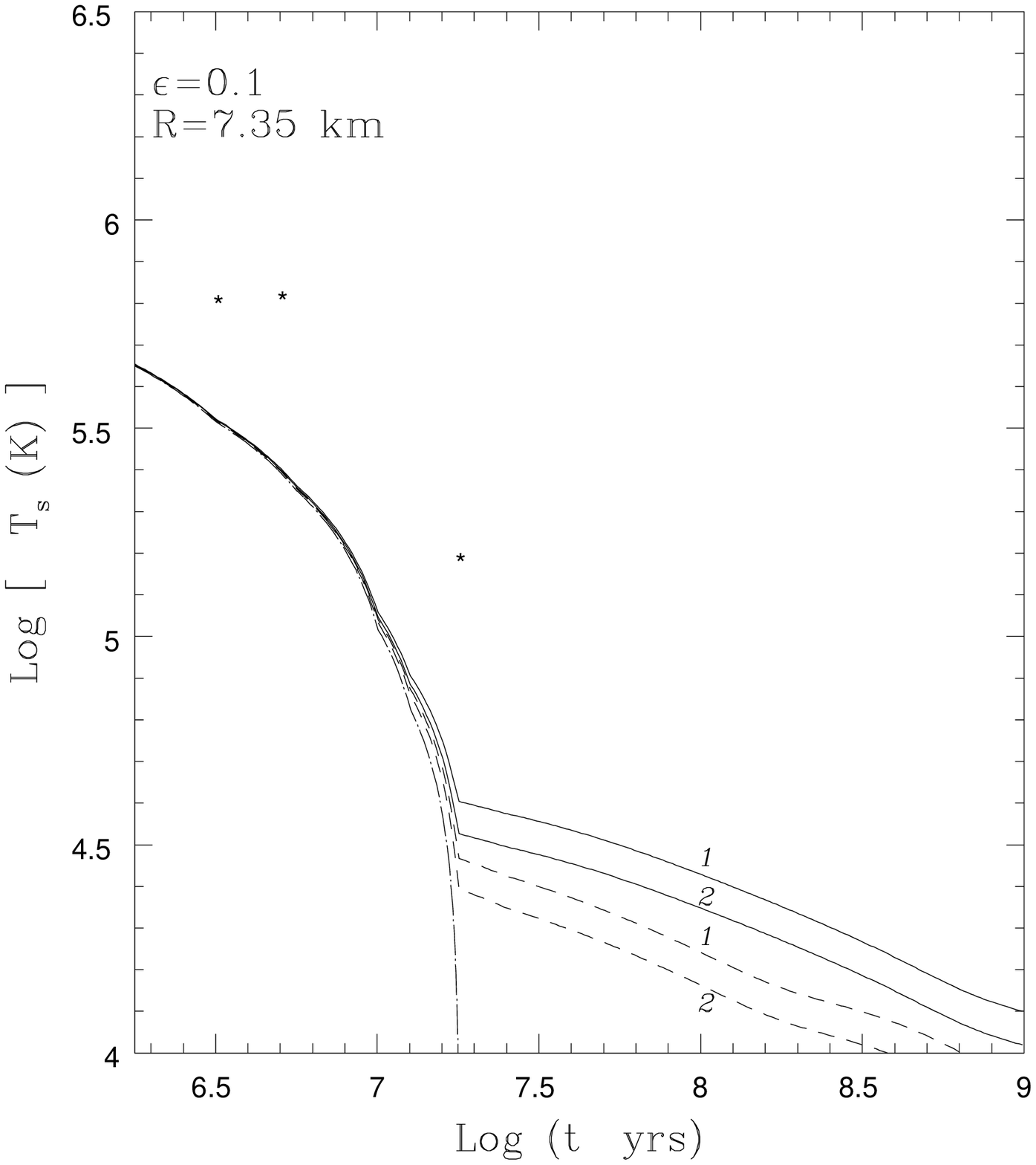]{Same as figure 5 but with $\epsilon=0.1$. \label{fig6}}
 
\figcaption[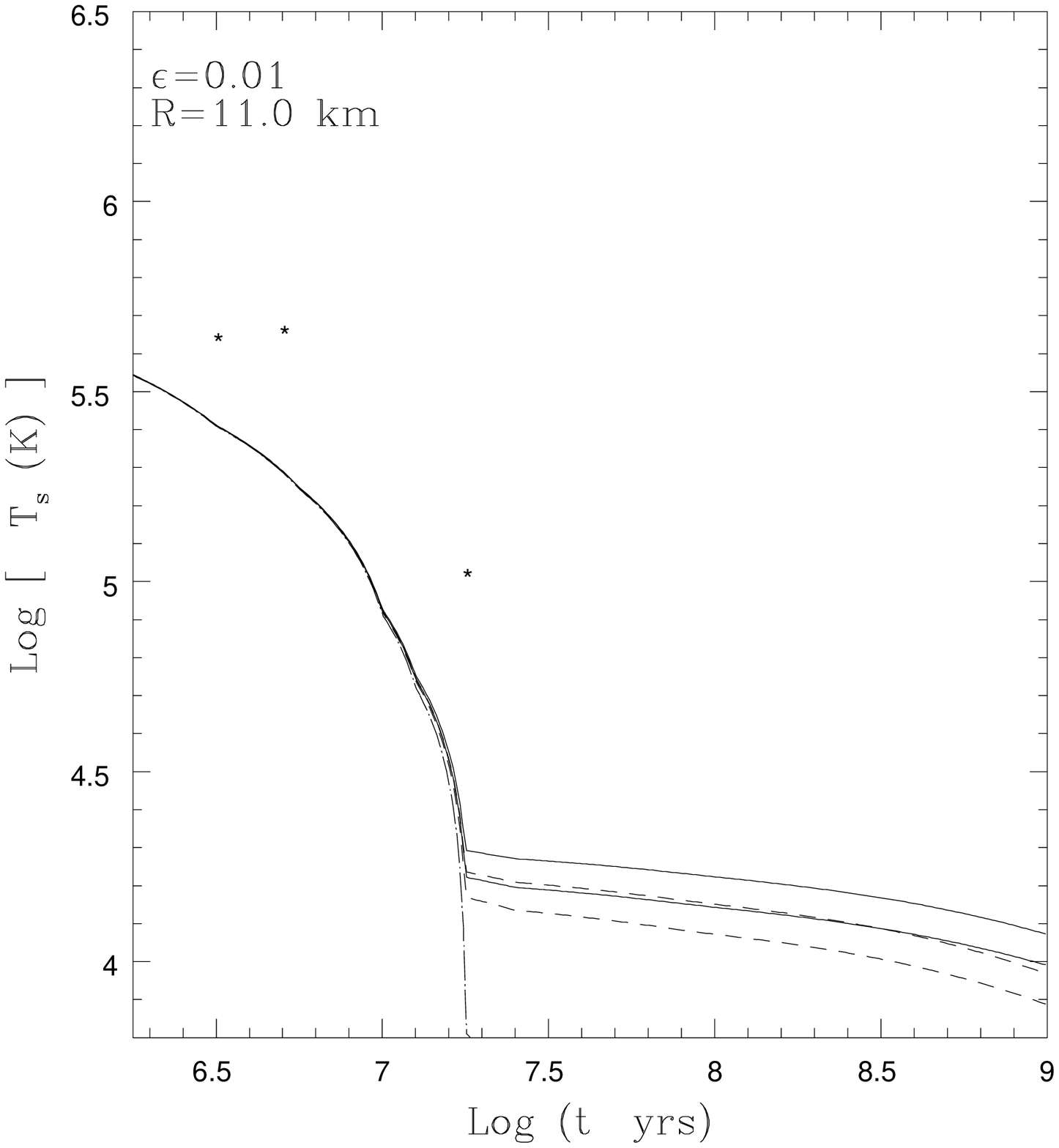] {Same as figure 5 but with the stellar radius 11 km. The upper
two curves (solid and broken lines) represent the results with $B_e=3.0\times
10^{13}$ G while the lower two curves represent the results with 
$B_e=1.5\times 10^{13}$ G. \label{fig7}}

\figcaption[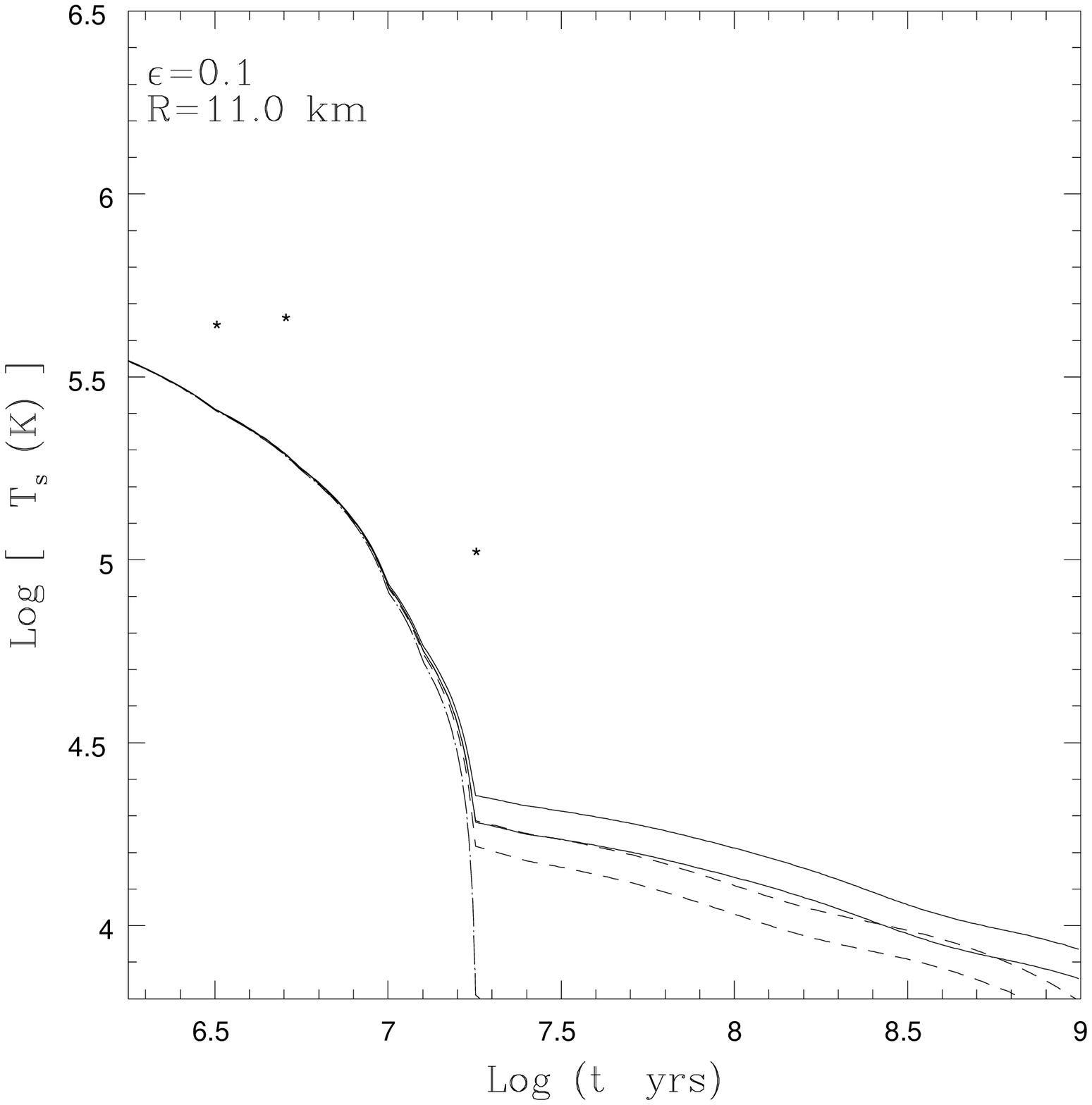] {Same as figure 7 but with $\epsilon=0.1$. The upper solid lines
and the upper broken lines represent the result with $B_e=3.0\times 10^{13}$ G
while the lower solid and broken lines represent that with $B_e=1.5\times
10^{13}$ G. \label{fig8}} 
\end{document}